# Magneto-thermal phenomena in bulk high temperature superconductors subjected to applied AC magnetic fields

P Vanderbemden[1], P Laurent[1], J-F Fagnard[2], M Ausloos[3], N Hari Babu[4] and D A Cardwell[5]

(1) SUPRATECS and Department of Electrical Engineering and Computer Science B28, Sart-Tilman, B-4000 Liège, Belgium

(2) SUPRATECS, Royal Military Academy of Belgium, Avenue de la Renaissance, B-1000 Brussels, Belgium

(3) SUPRATECS and Department of Physics B5, Sart-Tilman, B-4000 Liège, Belgium

(4) Brunel Centre for Advanced Solidification Technology (BCAST), Brunel University, West London, UB8 3PH, UK

(5) Bulk Superconductivity Group, Engineering Department, University of Cambridge, Cambridge, CB2 1PZ, UK

E-mail : Philippe.Vanderbemden@ulg.ac.be

**Abstract.** In the present work we study, both theoretically and experimentally, the temperature increase in a bulk high-temperature superconductor subjected to applied AC magnetic fields of large amplitude. We calculate analytically the equilibrium temperatures of the bulk sample as a function of the experimental parameters using a simple critical-state model for an infinitely long type-II superconducting slab or cylinder. The results show the existence of a limit heat transfer coefficient ($AU_{lim}$) separating two thermal regimes with different characteristics. The theoretical analysis predicts a "forbidden" temperature window within which the temperature of the superconductor can never stabilize when the heat transfer coefficient is small. In addition, we determine an analytical expression of two threshold fields $H_{tr1}$ and $H_{tr2}$ characterizing the importance of magneto-thermal effects and show that a thermal runaway always occurs when the field amplitude is larger than $H_{tr2}$. The theoretical predictions of the temperature evolution of the bulk sample during a self-heating process agree well with the experimental data. The simple analytical study presented in this paper enables order of magnitude thermal effects to be estimated for simple superconductor geometries under applied AC magnetic fields and can be used to predict the influence of experimental parameters on the self-heating characteristics of bulk type-II superconductors.

## 1. Introduction

The losses associated with the motion of vortices when type-II superconducting materials are penetrated by a variable magnetic field may give rise to a significant temperature increase and a degradation of the superconducting properties [1-3]. This potential problem is usually addressed in coated conductor or tape applications by sub-dividing the superconductor into "filaments" surrounded by a metallic matrix, which reduces significantly losses in the superconductor [4]. In addition, the metal matrix in the vicinity of the superconductor serves both to carry away heat generated locally and to distribute current between filaments if an individual filament is damaged [5]. A sub-division of the material, however, is not relevant to large, single grain bulk melt-processed superconductors, which are fabricated usually in disc or cylindrical geometries, for a variety of permanent magnet-like applications [6], such as magnetic bearings [7] and high power density rotating machines [8-10] due to their ability to trap large magnetic fields [11,12]. In this case, the material is likely to experience transient or periodic variations of the applied magnetic field that are due, for example, to vibrations, lateral movement or irregular magnetization of a permanent magnet that may interact with the superconductor [13,14]. Although the most severe damages arise when the time-varying field perturbations are perpendicular to the initial magnetization [15-17], the vortex motion in parallel configuration may also cause large hysteresis losses and possibly considerable self-heating which, in turn, may have a detrimental effect on the initial trapped flux of the bulk superconductor [18-28].

The problems of the temperature rise within bulk superconductors subjected to a variable magnetic field $H(t)$ may be classified as a function of the time-dependence of $H(t)$ as follows:

(i) Pulsed field magnetization (PFM), involving extremely large sweep rates (~ 100 T/s) during short times (~ 1-100 ms);
(ii) Traditional magnetization procedures, either during field cooled (FC) or after a zero field cooled (ZFC) magnetization process (i.e. at low sweep rates ranging typically between 1 and 10 mT/s);
(iii) AC magnetic fields, mainly at the frequency of the power line (50 or 60 Hz) and amplitudes up to ~ 100 mT.

The first situation (pulse field magnetization) leads usually to the most severe effects of self-heating, as demonstrated by experimental studies of bulk melt-processed materials [29-30]. For example, temperature rises of up to 30 K have been observed, depending on the initial temperature, the amplitude and duration of the pulsed field during the magnetization of bulk YBCO [29]. Although much smaller field sweep rates are involved, the second situation (magnetization during a FC or after a ZFC procedure), can induce temperature rises up to ~ 7 K [19]. These results indicate clearly that the generation of heat is a crucial parameter when estimating the field-trapping ability of high performance, bulk superconducting magnets. In the third category of experiments, the effects of the application of an AC magnetic field on a bulk melt-processed YBCO sample can lead to a significant temperature increase after a few minutes and, if the bulk material is magnetized permanently prior to application of the AC field, the losses can result in a severe decay, and even a complete collapse, of the trapped magnetic moment [22,23].

The temperature rise of a bulk superconductor subjected to an applied AC magnetic field, as usually observed in experiments [24,25] is shown schematically in Fig. 1. The initial temperature of the sample cooled cryogenically using liquid nitrogen at atmospheric pressure is represented by $T_0$ (= 77.4 K). The AC magnetic field $H(t) = H_m \sin(\omega t)$ is applied at $t = 0$. At low field amplitudes $H_m$, the sample temperature increases and reaches a steady-state value that is function of $H_m$ but smaller than the superconducting critical temperature $T_c$. Thermal runaway occurs on increasing the field amplitude further that results in a sharp temperature increase up to an equilibrium value close to $T_c$. The experimentally observed temperature rise of a bulk sample subjected to AC fields of varying amplitude can be described successfully by a simple analysis based on the Bean model [31] applied to a type-II superconducting cylinder assuming a linear temperature dependence of the critical current density $J_c(T)$ [22,32] and a constant convective heat transfer coefficient between the sample and the coolant [24]. In such studies performed, however, the theoretical $T(t)$ curves are obtained for a given set of material parameters and heat transfer characteristics that are representative of the superconducting sample under study. These parameters are the radius $a$ and volume $V$ of the cylindrical sample, the superconducting critical temperature $T_c$, coolant temperature $T_0$, initial critical current density $J_c(T_0)$, convective heat transfer coefficient $U$, contact area between the sample and the coolant $A$, heat capacity $c_p$, field amplitude $H_m$ and frequency of the applied field $f$. Values for each of these parameters can, of course, be modified in the modelling but the set of $T(t)$ curves generated consequently does not allow the most relevant parameters affecting the thermal behaviour of the sample either to be identified or to understand how they interact with each other.

The objective of the present work is to investigate further the mechanisms of self-heating of a bulk superconductor subjected to an AC field by determining *analytical* expressions for the sample temperature. The purpose of the investigation is to clarify the exact role of the different parameters indicated above. More specifically, this study will address the following issues:

(i) Determination of the exact steady-state (equilibrium) temperature of the sample as a function of the experimental parameters;

(ii) Investigation of whether the temperature behaviour illustrated schematically in Fig. 1 is always observed when a superconductor is subject to an AC field and whether is it possible to determine an analytical expression of the field amplitude separating the low-field (final temperature $<< T_c$) and high-field (final temperature $\sim T_c$) regimes;

(iii) Establish how the temperature behaviour is modified when the initial sample temperature differs from that of the coolant.

The paper is organized as follows. In Section 2, we first investigate the magneto-thermal effects arising in an infinite slab subjected to an AC magnetic field and then extend the results to the case of an infinite cylinder. In both cases the field is applied parallel to the long direction of the superconductor. The calculations are performed within the framework of the one-dimensional Bean model, i.e. finite-size effects and magnetic relaxation are neglected. Section 3 describes the experimental details of the temperature measurements carried out on a bulk melt-processed sample subjected to AC magnetic fields of large amplitude. The theoretical results are analyzed in Section 4 and compared with the experimental data. The conclusions are presented in Section 5.

## 2. Theory

*2.1. Magneto-thermal effects for an infinitely long superconductor*

In this section we consider an infinitely long type-II superconductor (infinite slab of thickness 2*a* or infinite cylinder of diameter 2*a*) and determine the magneto-thermal effects resulting from an AC magnetic field $H(t) = H_m \sin(2\pi f\, t)$ applied parallel to the long direction of the sample. The superconductor interacts with the cryogenic environment through a heat transfer coefficient $U$ and is assumed to be characterized by a thermal conductivity $\kappa$ much larger than the ($U\,a$) product. In such a case the dimensionless Biot number $Bi = U\,a\,/\,\kappa$ is much smaller than unity and the sample temperature $T$ can be taken as uniform [28]. The analysis is based on the one-dimensional Bean model for a field-independent critical current density $J_c$ [31]. In both cases (slab or cylinder), the full-penetration field $H_p(T)$ at a given temperature $T$ is given by;

$$H_p(T) = J_c(T)\,a\,. \tag{1}$$

The hysteresis losses caused by the AC field are usually expressed as a function of the dimensionless parameter $\beta$ (normalized magnetic field amplitude) defined by;

$$\beta = \frac{H_m}{H_p}\,. \tag{2}$$

The superconductor is partially penetrated for $\beta < 1$ and fully penetrated for $\beta > 1$. The AC losses $Q_{gen}$ [expressed in W] generated within a superconductor of volume $V$ for the various approximations are given by [3,4]:

In the infinite slab approximation:

$$Q_{gen} = 2\,\mu_0\,f\,V\,H_m^2\left(\frac{\beta}{3}\right) \qquad \text{for } \beta \leq 1 \tag{3}$$

$$Q_{gen} = 2\,\mu_0\,f\,V\,H_m^2\left(\frac{1}{\beta} - \frac{2}{3\beta^2}\right) \qquad \text{for } \beta \geq 1 \tag{4}$$

In the infinite cylinder approximation:

$$Q_{gen} = 2\,\mu_0\,f\,V\,H_m^2\left(\frac{2\beta}{3} - \frac{\beta^2}{3}\right) \qquad \text{for } \beta \leq 1 \tag{5}$$

$$Q_{gen} = 2\,\mu_0\,f\,V\,H_m^2\left(\frac{2}{3\beta} - \frac{1}{3\beta^2}\right) \qquad \text{for } \beta \geq 1 \tag{6}$$

In this analysis, the temperature-dependence of $J_c$ is assumed to be linear [22,32], although more refined models can be used [33,34]. The $J_c(T)$ law is given by

$$J_c(T) = J_{c0}\left(\frac{T_c - T}{T_c - T_0}\right), \qquad (7)$$

where $T_0$ is the coolant temperature (i.e. 77.4 K), $T_c$ is the critical temperature and $J_{c0}$ is the critical current density at $T = T_0$. The AC losses $Q_{gen}$ become a function of temperature when the temperature-dependence of $J_c$ in Eq. (1)-(6) is introduced to the formulation, as illustrated in Fig. 2 (black curves). It can be seen that the losses $Q_{gen}(T)$ between $T_0$ and $T_c$ exhibit a well-defined maximum, above which they decrease down to 0 at $T = T_c$. The AC field amplitude $H_m$ in Fig. 2, is assumed to be smaller than the full-penetration field $H_{p0}$ at $T_0$, i.e.;

$$H_{p0} = J_{c0}\, a. \qquad (8)$$

It should be noted that the partially penetrated ($\beta < 1$) and the fully penetrated ($\beta > 1$) regimes for an infinite slab correspond to parts located on the left and on the right hand-sides of the *inflection point* appearing in $Q_{gen}(T)$ in Fig. 2, whereas the partially and fully penetrated regimes for the infinite cylinder are delimited by the *maximum* of $Q_{gen}(T)$, i.e. the losses are maximum for $\beta = 1$.

The AC losses in the superconductor are associated with heat transfer through the external surface of the sample. Previous works have shown the importance of thermal boundary conditions [35]. For simplicity, we assume that the heat transfer rate $Q_{out}$ between the superconductor and the cryogenic environment (at temperature $T_0$) is given by [22,24]

$$Q_{out} = AU(T - T_0), \qquad (9)$$

where $A$ represents the contact area between the sample and the coolant [units: m²], $U$ denotes the convective heat transfer coefficient [units: Wm$^{-2}$K$^{-1}$], and $T$ is the sample temperature. The ($AU$) product will be considered as a single parameter [units: WK$^{-1}$] in the following analysis, and considered to characterize the heat transfer between the sample and the cryogenic fluid. According to Eq. (9), the heat transfer rate $Q_{out}(T)$ is a linear function of temperature, and crosses the horizontal axis at $T = T_0$, as shown in Fig. 2. The thermal behaviour of the superconductor is then simply predicted by the values of $Q_{gen}$ and $Q_{out}$, respectively; i.e. if $Q_{gen} > Q_{out}$ the sample temperature $T$ increases, if $Q_{gen} < Q_{out}$, $T$ decreases and the thermal equilibrium corresponds to $Q_{gen} = Q_{out}$. We now examine the various regimes of behaviour illustrated in Fig. 2.

The value of $AU$ (i.e. the slope of the $Q_{out}(T)$ line) in Fig. 2(a) is rather small and such that five different regimes are observed, depending on the amplitude of magnetic field $H_m$. At small field amplitudes ($H_m < H_{tr1}$), the $Q_{gen}$ and $Q_{out}$ curves intersect at some temperature $T_{inf}$, slightly above $T_0$. The temperature increases up to $T_{inf}$ (steady-state temperature) if the initial sample temperature is smaller than $T_{inf}$, and decreases down to $T_{inf}$ if the initial sample temperature exceeds $T_{inf}$. This crossing point corresponds thus to a *stable* thermal equilibrium, since a slight disturbance above this temperature corresponds to $Q_{gen} < Q_{out}$, i.e. a cooling of the sample down to the former temperature $T_{inf}$. Similarly, when temperature is slightly decreased below $T_{inf}$, the $Q_{gen} > Q_{out}$ condition causes the temperature to return to its original value. When the AC field amplitude increases, a second ($T_{sup}$, for $H_m = H_{tr1}$) and a third ($T_{med}$, for $H_{tr1} < H_m < H_{tr2}$) intersection points occur. It can be readily checked that the temperature $T_{sup}$ corresponds to a *stable* equilibrium temperature whereas the intersection point at $T = T_{med}$ is *unstable* and cannot be considered as equilibrium temperature for the system. At medium field amplitudes ($H_{tr1} \leq H_m \leq H_{tr2}$), the sample temperature can thus be stabilized at *two* points $T_{inf}$ and $T_{sup}$.

In the common case where the initial sample temperature is equal to that of the cryogenic fluid (i.e. $T = T_0$), the AC losses in the superconductor cause the temperature to increase up to $T_{inf}$, which corresponds to a stable equilibrium. Consequently, the only observed steady-state temperature in this regime is $T_{inf}$, provided the sample is at an initial temperature of $T_0$. The lower equilibrium temperature disappears (at $H_m = H_{tr2}$) on increasing the magnetic field further, and only one equilibrium temperature remains, i.e. $T = T_{sup}$ at for field amplitudes larger than $H_{tr2}$.

The different regimes displayed in Fig. 2(a) are separated by two threshold magnetic fields $H_{tr1}$ and $H_{tr2}$ that are function of the experimental parameters. However, two situations occur when the superconductor is placed initially at $T_0$, i.e.;

(i)   for $H_m < H_{tr2}$, the temperature increases up to $T_{inf}$
(ii)  for $H_m > H_{tr2}$, the temperature increases up to $T_{sup}$

As a consequence, the relevant magnetic field defining the transition between the two equilibrium temperatures is the upper threshold field $H_{tr2}$. As can be observed in Fig. 2(a), a clear discontinuity arises at $H_m = H_{tr2}$, since the highest possible $T_{inf}$ is smaller than the smallest $T_{sup}$. This indicates that the final equilibrium temperature of the superconductor cannot occur within the interval $T_{inf}(H_{tr2}) - T_{sup}(H_{tr2})$; this temperature interval is delimited by the vertical dashed lines shown in Fig. 2(a).

The "forbidden" temperature range mentioned above only exists in the case where the heat transfer coefficient, $AU$, is small enough to give rise to the different regimes shown in Fig. 2(a). For large $AU$ values (Fig. 2(c)), there is only *one* intersection point between the $Q_{gen}(T)$ curve the $Q_{out}(T)$ straight line, whatever the amplitude of AC field $H_m$. In this case, the final equilibrium temperature of the superconductor is a *continuous* and increasing function of $H_m$. This behaviour allows us to define a *limit heat transfer coefficient*, $AU_{lim}$, separating both situations; the corresponding curves are shown in Fig. 2(b). For $AU = AU_{lim}$, there is a given amplitude of AC field (referred to as "$H_{lim}$") for which the intersection point between $Q_{gen}$ and $Q_{out}$ occurs precisely at the inflection point of $Q_{gen}(T)$. This simple property can be used to determine the analytical expressions of both $AU_{lim}$ and $H_{lim}$, as will be described below.

Having established the different scenario in a general manner, we now turn to determining the analytical expressions for the threshold parameters $AU_{lim}$, $H_{lim}$, $H_{tr1}$ and $H_{tr2}$, together with the different possible equilibrium temperatures. We examine successively the case of an infinite slab (Section 2.2) and of an infinite cylinder (Section 2.3).

*2.2. Threshold parameters and equilibrium temperatures for an infinite slab*

In this section, we use the analytical expressions of the losses for an infinite slab [Eq. (3)-(4)] in order to determine the different intersection points and regimes illustrated in Fig. (2).

First, the limit field $H_{lim}$ corresponds to the field amplitude at which the tangent line at the inflection point of the $Q_{gen}(T)$ curve cuts the horizontal axis at $T = T_0$. Inserting (7)-(8) into Eq. (3) gives;

$$H_{lim} = \frac{1}{2}(J_{c0}\, a) = \frac{H_{p0}}{2}. \tag{10}$$

The limit heat transfer coefficient, $AU_{lim}$, corresponds to the slope of the tangent of $Q_{gen}(T)$ at its inflection point, i.e.;

$$AU_{lim} = \left(\frac{dQ_{gen}}{dT}\right)_{H_m = H_{lim}} = \left(\frac{1}{3}\right)\frac{\mu_0\, f\, V\, H_{p0}^2}{T_c - T_0} \tag{11}$$

It is necessary to calculate the intersection points between the $Q_{gen}(T)$ and $Q_{out}(T)$ curves when $AU < AU_{lim}$ to determine the threshold fields $H_{tr1}$ and $H_{tr2}$. We introduce dimensionless temperature ($\tilde{t}$), magnetic field ($x$) and heat transfer coefficient ($\alpha$) given by

$$\tilde{t} = \frac{T - T_0}{T_c - T_0}, \tag{12}$$

$$x = \frac{H_m}{H_{lim}}, \tag{13}$$

$$\alpha = \frac{AU}{AU_{lim}}. \tag{14}$$

It is also of interest to notice that the normalized heat transfer coefficient $\alpha$ can be expressed as follows;

$$\alpha = \frac{2\,AU\,(T_c - T_0)}{\left(\frac{2}{3}\right)\mu_0\,f\,V\,H_{p0}^2} = 2\,\frac{Q_{out}(T_c)}{Q_{gen}(H_p)}, \tag{15}$$

where $Q_{out}(T_c)$ is heat transferred when the sample temperature is equal to $T_c$ and $Q_{gen}(H_p)$ denotes the losses at a magnetic field amplitude equal the full-penetration field.

The equations corresponding to the equality $Q_{gen}(T) = Q_{out}(T)$ are given by

$$4\alpha\,\tilde{t}^{\,2} - 4\alpha\,\tilde{t} + x^3 = 0 \qquad \text{for } \tilde{t} \leq 1 - (x/2) \tag{16}$$

$$4(1-\tilde{t})^2 - (\alpha + 3x)(1-\tilde{t}) + \alpha = 0 \qquad \text{for } \tilde{t} \geq 1 - (x/2) \tag{17}$$

The normalized lower threshold field $x_{tr1} = (H_{tr1}/H_{lim})$ corresponds to the value of $x$ for which Eq. (17) has a double root and is given by

$$x_{tr1} = \frac{H_{tr1}}{H_{lim}} = \frac{1}{3}\left(-\alpha + 4\sqrt{\alpha}\right). \tag{18}$$

Using the expression of $H_{lim}$ [Eq. (10)], the lower threshold field $H_{tr1}$ reads

$$H_{tr1} = -\frac{AU\,(T_c - T_0)}{2\,\mu_0 f\,V\,H_{p0}} + 2\sqrt{\frac{AU\,(T_c - T_0)}{3\,\mu_0 f\,V}}. \tag{19}$$

Similarly, the $x$ value for which Eq. (16) has a double root corresponds to the normalized upper threshold field $x_{tr2} = (H_{tr2}/H_{lim})$ given by;

$$x_{tr2} = \frac{H_{tr2}}{H_{lim}} = \alpha^{1/3}, \tag{20}$$

which can be rewritten using the expression of $H_{lim}$, i.e.;

$$H_{tr2} = \left[ \frac{3\,AU\,(T_c - T_0)H_{p0}}{8\,\mu_0 f\,V} \right]^{1/3}. \tag{21}$$

Remarkably, equations (18) and (20) show that the normalized threshold fields can be expressed as a function of a *single* parameter $\alpha = AU / AU_{lim}$. It can be checked readily that both normalized threshold fields are equal to 1 when $\alpha = 1$.

Now we turn to the determination of the lower and upper equilibrium temperatures $T_{inf}$ and $T_{sup}$. The resolution of Eq. (16) in normalized units gives directly;

$$\tilde{t}_{inf} = \frac{T_{inf} - T_0}{T_c - T_0} = \frac{1}{2}\left[ 1 - \sqrt{1 - \frac{x^3}{\alpha}} \right]. \tag{22}$$

Equation (22) shows clearly that the normalized lower equilibrium temperature ($\tilde{t}_{inf}$) is always smaller than or equal to 1/2; $\tilde{t}_{inf}$ it is equal to 1/2 [i.e. $T_{inf} = (T_0 + T_c)/2$] when the field amplitude equals the upper threshold field, i.e. when $x = \alpha^{1/3}$. At small field amplitudes $x \ll \alpha$ ($H_m \ll H_{tr2}$), the lower equilibrium temperature increases as a cubic function of the applied field amplitude, i.e.;

$$\tilde{t}_{inf} \approx \frac{x^3}{4\alpha}. \tag{23}$$

In a similar manner, the resolution of Eq. (17) gives the normalized upper equilibrium temperature $\tilde{t}_{sup}$:

$$1-\tilde{t}_{sup} = \frac{T_c - T_{sup}}{T_c - T_0} = \frac{\alpha}{8}\left(1+\frac{3x}{\alpha}\right)\left[1-\sqrt{1-\frac{16}{\alpha\left(1+\frac{3x}{\alpha}\right)^2}}\right]. \quad (24)$$

At large field amplitudes $x \gg \alpha$ ($H_m \gg H_{tr2}$), Eq. (24) can be approximated by

$$1-\tilde{t}_{sup} = \frac{T_c - T_{sup}}{T_c - T_0} \approx \frac{\alpha}{3x}. \quad (25)$$

Equation (25) shows that, at large field amplitudes, the sample temperature follows a hyperbolic function of the field amplitude and approaches $T_c$ asymptotically.

*2.3. Threshold parameters and equilibrium temperatures for an infinite cylinder*

In this section, we examine the case of an infinite cylinder (radius *a*) and determine the same parameters to those calculated above for the infinite slab, i.e. the limit heat transfer coefficient and magnetic field $AU_{lim}$ and $H_{lim}$, the two threshold fields $H_{tr1}$ and $H_{tr2}$ and the equilibrium temperatures $T_{inf}$ and $T_{sup}$. Using the same procedure as that used for the infinite slab, one finds;

$$H_{lim} = \frac{2}{9}(J_{c0}\, a) = \frac{2}{9}H_{p0}. \quad (26)$$

$$AU_{lim} = \left(\frac{dQ_{gen}}{dT}\right)_{H_m = H_{lim}} = \left(\frac{32}{729}\right)\frac{\mu_0\, f\, V\, H_{p0}^2}{T_c - T_0}. \quad (27)$$

The values of $H_{lim}$ and $AU_{lim}$ for the infinite cylinder [Eq. (26)-(27)] differ from their counterparts determined for the infinite slab [Eq. (10)-(11)] by a numerical factor but they exhibit the same dependence with respect with the experimental parameters *f*, *V*, $H_{p0}$, $T_c$ and $T_0$.

The equations corresponding to the equality $Q_{gen}(T) = Q_{out}(T)$ for an infinite cylinder are given by

$$\tilde{t}^3 - 2\tilde{t}^2 + \left(1 + \frac{x^3}{3\alpha}\right)\tilde{t} + \frac{x^3}{3\alpha}\left(\frac{x}{9} - 1\right) = 0, \qquad \text{for } \tilde{t} \leq 1 - (2x/9) \qquad (28)$$

$$\frac{243}{16\alpha}(1-\tilde{t})^2 - \left(1 + \frac{27x}{4\alpha}\right)(1-\tilde{t}) + 1 = 0, \qquad \text{for } \tilde{t} \geq 1 - (2x/9) \qquad (29)$$

using the dimensionless variables $\tilde{t}$, $x$ and $\alpha$ defined by Eqs. (12)-(14). The normalized threshold fields $x_{tr1} = (H_{tr1} / H_{lim})$ and $x_{tr2} = (H_{tr2} / H_{lim})$ correspond to the field amplitudes values for which Eq. (28) has a double root. The general condition for which a cubic equation $a\tilde{t}^3 + b\tilde{t}^2 + c\tilde{t} + d = 0$ has a double root is given by [36];

$$-4b^3 d + b^2 c^2 - 4ac^3 + 18abcd - 27 a^2 d^2 = 0. \qquad (30)$$

Applying this condition to Eq. (28) leads to;

$$4x^5 + \alpha x^4 - 6\alpha x^3 - 3\alpha x^2 + 4\alpha^2 = 0. \qquad (31)$$

This polynomial equation has two real (positive) roots for $\alpha < 1$ that correspond to the normalized threshold fields $x_{tr1}$ and $x_{tr2}$. The threshold fields for $\alpha > 1$ do not have physical meaning and Eq. (31) has no real positive root. In the particular case of $\alpha = 1$, Eq. (31) has a double root $x = 1$, which corresponds to $H_{tr1} = H_{tr2} = H_{lim}$. In the general case, Eq. (31) is not soluble analytically and the threshold fields have to be computed numerically. For small values of $\alpha$, however, approximate analytical solutions can be found. By analogy with the exact expressions obtained for the infinite slab [Eq. (18) and (20)], we assume that $x_{tr1}$ and $x_{tr2}$ for the cylinder follow respectively a $\alpha^{1/2}$ and a $\alpha^{1/3}$ dependence for $\alpha \ll 1$. Hence approximate expressions of the lower and upper threshold fields $H_{tr1}$ and $H_{tr2}$ are given by;

$$x_{tr1} = \frac{H_{tr1}}{H_{lim}} \approx \left(\frac{4}{3}\right)^{1/2} \alpha^{1/2}, \qquad (32)$$

$$x_{tr2} = \frac{H_{tr2}}{H_{lim}} \approx \left(\frac{3}{4}\right)^{1/3} \alpha^{1/3}. \qquad (33)$$

Now we turn to the determination of the lower and upper equilibrium temperatures, corresponding to the roots of Eq. (28) and (29), respectively. The analytical expression of lower equilibrium temperature $\tilde{t}_{inf}$ is not reproduced in this paper but can be obtained from standard formulas for the roots of a cubic function [36]. When the field amplitude equals the upper threshold field ($x = x_{tr2}$), Eq. (28) admits a double root given by

$$\tilde{t}_{inf}(x = x_{tr2}) = \left[\frac{2}{3} - \frac{1}{3}\sqrt{1 - \frac{(x_{tr2})^3}{\alpha}}\right]. \qquad (34)$$

At small heat transfer coefficients ($\alpha \ll 1$), Eq. (33) and (34) show that the lower equilibrium temperature at the upper threshold field is equal to $\tilde{t}_{inf} = 1/2$ [i.e. $T_{inf} = (T_0 + T_c)/2$], as is the case for the infinite slab. In the particular case $\alpha = 1$, corresponding to $x_{tr2} = 1$, the lower equilibrium temperature at the upper threshold field is equal to $\tilde{t}_{inf} = 2/3$ [i.e. $T_{inf} = (T_0 + 2T_c)/3$]. At small field amplitudes, the lower equilibrium temperature increases as a cubic function of the field amplitude, i.e.;

$$\tilde{t}_{inf} \approx \frac{x^3}{3\alpha}. \qquad (35)$$

The analytical expression of the upper equilibrium temperature is obtained by solving Eq. (29), i.e;.

$$1 - \tilde{t}_{sup} = \frac{T_c - T_{sup}}{T_c - T_0} = \frac{8\alpha}{243}\left(1 + \frac{27x}{4\alpha}\right)\left[1 - \sqrt{1 - \frac{243}{4\alpha\left(1 + \frac{27x}{4\alpha}\right)^2}}\right]. \qquad (36)$$

At large field amplitudes, $x \gg \alpha$ ($H_m \gg H_{tr2}$), Eq. (36) can be approximated by

$$1 - \tilde{t}_{sup} = \frac{T_c - T_{sup}}{T_c - T_0} \approx \frac{4\alpha}{27x}. \tag{37}$$

In summary, the equations above allow the characteristics of the thermal behaviour of an infinite type-II superconductor subjected to an AC field to be determined. Before analyzing the results, we describe briefly the experimental procedure used to measure magneto-thermal effects in bulk melt-processed superconductors.

3. Experiment

Bulk, melt-processed single domains of YBCO, consisting of a superconducting $YBa_2Cu_3O_{7-\delta}$ (Y-123) matrix with discrete $Y_2BaCuO_5$ (Y-211) inclusions, were fabricated by top seeded melt growth (TSMG), as described in Refs. [37-40]. The present study focuses on a single grain sample (PL1) consisting of a single domain pellet of 30 mm diameter and 12 mm thickness. The material is characterized by a critical temperature $T_c \sim 91.6$ K. A Hall probe-mapping experiment carried out above the top-surface of the sample reveals a critical current density $J_c$ of $\sim 10^3$ A/cm² at 77.4 K. We note that the value of critical current density for this sample is below the average level that can be obtained using the TSMG technique [37]. A sample of medium quality with uniform superconducting properties was selected specifically for the magneto-thermal measurements reported here because, in the partially penetrated state, the AC losses are (to a first approximation) inversely proportional to the critical current density $J_c$ [see Eq. (1)-(3)]. A relatively "low" $J_c$ ensures that sufficient losses will be generated within the sample at 77.4 K for experimentally accessible values of the applied magnetic field amplitude $H_m$.

Magneto-thermal measurements on this YBCO sample were carried out in a bespoke AC susceptometer designed for the characterization of large superconducting samples (up to 32 mm diameter). The apparatus is described in details in Ref. [41]; the experimental chamber is schematically illustrated in Fig. 3(a). In this system the superconductor is placed in a sample chamber connected to a vacuum rotary pump, enabling a medium vacuum ($p \sim 2 - 9 \cdot 10^{-2}$ mbar at room temperature) to be achieved. The sample chamber walls are made of ultra-high molecular weight polyethylene (PE-UHMW). The sample chamber is immersed directly in liquid nitrogen and no helium gas is used in the measurement. In order to minimize the thermal contact between the superconductor and the bottom of the sample chamber, the sample is deposited against an alumina disk placed on 3 glass spheres (2.4 mm diameter) located at the vertices of a triangle. This configuration ensures a small, but reproducible, heat flux rate out of the sample. The temperature of the sample is monitored using three type-E thermocouples (chromel-constantan) attached to the top surface at various distances from the centre.

The melt-processed sample is initially cooled to 77.4 K in zero applied field (ZFC procedure) in all the experiments performed here. This differs from previous studies in which the sample is magnetized permanently prior to the application of the AC field [22-24]. The alternating magnetic field is applied parallel to the $c$-axis of the sample and the evolution of temperature with time recorded either with Keithey 2001 voltmeters (at small $dT/dt$, typically < 5 K /min) or with AD623 instrumentation amplifiers (when $dT/dt$ > 5 K /min) using a PCI 6221 DAQ board from National Instruments.

The sample is cooled initially to the lowest possible temperature (i.e. $T_0$ = 77.4 K) in order to determine precisely the heat flux rate out of the superconductor in the present experimental system. An AC magnetic field $H(t) = H_m \sin(2\pi f t)$ is then applied and magnetic AC losses $Q_{gen}(H_m)$ recorded by the susceptometer. The losses induce self-heating which causes the temperature to increase to some equilibrium value $T_e(H_m)$ that is measured by the thermocouples attached to the sample. The equilibrium condition ensures that the magnetic losses and the heat flux rate out of the sample are equal, i.e. $Q_{gen}(H_m) = Q_{out} = AU[T_e(H_m) - T_0]$. Measurements of $Q_{gen}(H_m)$ and $T_e(H_m)$ at several field amplitudes $H_m$ allow the heat transfer coefficient $AU$ to be determined by linear regression, as illustrated in Fig. 3(b). Such a procedure leads to $AU = 4.94 \times 10^{-3}$ W/K, i.e. a convective coefficient $U$ = 1.94 W/m²K, which is 3-4 orders of magnitude lower than the typical values of the pool boiling heat transfer coefficient from YBCO to liquid nitrogen [42,43]. The quasi-linear behaviour observed between $Q_{gen}$ and $T_e$ justifies, *a posteriori*, the validity of the linear $Q_{out} = AU (T-T_0)$ law [Eq. (9)]. The very low scatter of the data points (obtained for different measurement runs) also emphasizes that the experimental system is characterized by a reproducible heat transfer coefficient. This is direct evidence that the apparatus is fully adequate for studying magneto-thermal effects in bulk high-temperature superconductors.

## 4. Results and discussion

*4.1. Threshold fields as a function of the heat transfer coefficient*

Figure 4 shows the analytically calculated threshold fields $H_{tr1}$ and $H_{tr2}$ as a function of the heat transfer coefficient $AU$. A log-log scale is used to emphasize the power law behaviour. The heat transfer coefficient $AU$ is normalized to the limit value $AU_{lim}$, whereas both threshold fields are normalized to the limit field $H_{lim}$ defined in Sect. 2.1.

We first consider the threshold fields for the infinite slab, represented by plain lines in Fig. 4. According to Eq. (21), $H_{tr2}$ follows a $(AU)^{1/3}$ behaviour, as appears by a straight line in Fig. 4. At low values of $\alpha = AU/AU_{lim}$ ($\ll 1$), the threshold field $H_{tr1}$ follows roughly a power law with exponent 1/2, as is anticipated from Eq. (19). At $AU = AU_{lim}$, both threshold fields merge into one and their value is equal to $H_{lim}$, as predicted by the theory. For $AU > AU_{lim}$, the threshold fields do not have any physical meaning since only one self-heating regime occurs. The threshold field values for cylinder, obtained from the numerical resolution of Eq. (31) for several $\alpha$ values, are shown by symbols in Fig. 4. The approximations for $\alpha \ll 1$ [Eq. (32) and (33)] appear by dashed lines. As can be seen in Fig. 4, the threshold field values for the slab and cylinder geometries are very similar to each other when they are normalized to their "limit" counterparts, in spite of different numerical values of $AU_{lim}$ and $H_{lim}$. This suggests that the kind of approximation that can be used in practice to model the behaviour of a given sample that differs from the slab or cylinder geometry (e.g. a parallelepiped) is not crucial; both cylinder and slab approximations yield similar threshold fields.

The variation of $H_{tr2}$ with $AU$ is of practical relevance for a given experiment, since it allows directly the kind of magneto-thermal behaviour to be predicted, depending on the specific values of $H_m$ and $AU$ for the particular system. If the applied field amplitude $H_m$ is such that the ($H_m$, $AU$) point is located *below* the $H_{tr2}$ vs. $AU$ line, then self-heating effects can be considered to be small and the final equilibrium temperature is less than $(T_0 + T_c)/2$ for a slab, and less than $(T_0 + 2T_c)/3$ for a cylinder. When the $(H_m, AU)$ point is located *above* the $H_{tr2}$ vs. $AU$ line, however, magneto-thermal effects are significant and the equilibrium temperature is close to (but slightly smaller than) $T_c$. If the $(H_m, AU)$ point is located *on the right-hand side of* the $H_{tr2}$ vs. $AU$ line, i.e. for $AU > AU_{lim}$, the final sample temperature is an increasing continuous function of the field amplitude $H_m$. Note that, unless extremely high AC field amplitudes are used, self-heating effects in this regime are expected to be minor since the heat transfer coefficient, indicative the cooling efficiency, is assumed to be high.

*4.2. Equilibrium temperature of the superconductor as a function of AC field amplitude*

In this section we use the analytical expressions derived in Section 2 to characterize the magneto-thermal behaviour of an infinite type-II superconducting slab subjected to an AC magnetic field $H(t) = H_m \sin(2\pi f t)$. Fig. 5 shows the final temperature (thermal equilibrium) attained by a superconductor having a $T_c$ of 92 K when it is placed at an initial temperature $T_0 = 77.4$ K. The amplitude of magnetic field is normalized to the upper threshold field $H_{tr2}$ and three situations are considered, depending on the heat transfer coefficient $AU$ with respect to the limiting value $AU_{lim}$. The equilibrium temperature for $AU < AU_{lim}$, is shown to exhibit a discontinuity at $H_m = H_{tr2}$, as predicted in Section 2(a). This indicates a "forbidden" temperature range in which the superconductor temperature can never stabilize. The final temperature in the "low temperature range" ($H_m < H_{tr2}$) exhibits a quasi-cubic law behaviour, as predicted by Eq. (23). The final temperature in the "high temperature range" ($H_m > H_{tr2}$) approaches asymptotically the critical temperature $T_c$, as shown by Eq. (25). Strictly speaking, the theory predicts that the final equilibrium temperature is always *smaller* than $T_c$, even at large field amplitudes. This is consistent with experimental observations [26,44] of the existence of small, but finite, AC losses in this regime.

The "forbidden" temperature window apparent in Fig. 5(a) narrows as the heat transfer coefficient $AU$ increases and eventually vanishes at $AU = AU_{lim}$ (Fig. 5(b)). A continuous field-dependence of the equilibrium temperature is observed above $AU_{lim}$ (Fig. 5(c)). We note further that, for given a normalized field amplitude, the steady-state temperature is a decreasing function of the heat transfer coefficient, as expected intuitively.

*4.3. Comparison with experimental results*

Figure 6(a) shows the theoretical equilibrium temperature of an infinitely long cylindrical sample as a function of the AC field amplitude, using the experimental parameters given in Table 1. The heat transfer coefficient $AU$ is the value determined experimentally in Section 3 ($4.94 \times 10^{-3}$ W/K) and is smaller than the limit value $AU_{lim}$ determined from experimental parameters in Eq. (27), i.e. $AU_{lim} = 41.7 \times 10^{-3}$ W/K. The corresponding threshold field $\mu_0 H_{tr2}$ is equal to 19.8 mT. The data shown in Fig. 6(a), corresponding to $AU < AU_{lim}$, indicate the existence of two equilibrium regimes separated by the upper threshold field $H_{tr2}$, as well as the forbidden temperature range within which the sample temperature cannot stabilize.

Figure 6(b) shows the magnetic field dependence of the steady-state temperature of a bulk, melt-processed YBCO disc (sample PL1) subjected to an AC magnetic field. The experimental parameters are, again, those listed in Table 1. It can be seen that the experimental data show all the qualitative features predicted theoretically and which are evident in Fig. 6(b): (i) two equilibrium regimes are visible, separated clearly by a threshold field determined experimentally here to be 26.95 mT; (ii) the equilibrium temperature for $H_m < H_{tr2}$ increases with increasing AC field amplitude and follows closely a cubic law; (iii) the equilibrium temperature for $H_m > H_{tr2}$ is close to the sample critical temperature $T_c$. Remarkably, the experimental equilibrium temperature in this regime reveals an unambiguous correlation with the amplitude of the applied field, as shown by the data in the inset of Fig. 6(b). This behaviour is in good agreement with the theoretical predictions shown in Fig. 6(a) for the infinite cylinder. Note that from an experimental point of view, the temperature resolution achieved via the thermocouple readings is approximately 0.02 K, corresponding to a voltage resolution of ~ 0.5 µV. Such sensitivity explains the scatter of the data in the inset of Fig. 6(b) and underlines the great care required observe meaningfully magneto-thermal effects in which DC thermocouple voltages are recorded in the presence of a large (parasitic) AC magnetic field.

Although both theoretical and experimental results agree qualitatively, the experimental threshold field (26.95 mT) is found to differ from the theoretical prediction (19.8 mT for the infinite cylinder approximation) by a factor of 1.73. This feature is likely to be attributed to three factors. Firstly, the true geometry of the sample, i.e. a cylindrical disc of aspect ratio height/diameter = 0.4 differs from an infinite cylinder. The consequence of this is that the AC magnetic field penetration occurs not only from the lateral surface of the sample but also from its top and bottom circular surfaces, as is well established for thin or flat superconductors in perpendicular orientation [45-48]. It should be noted that the presence of surface in a superconducting cylinder of finite height is also beneficial in terms of cooling efficiency since it increases the contact area $A$ between the sample and the coolant. This latter point might be predominant in the present case since the experimental threshold field is larger than the theoretical value. The second point to be considered is that the theoretical predictions presented here are based on a critical current density $J_c$ value ($10^3$ A/cm²) extracted from the experimental magnetic field distribution measured above the top surface of the sample (Hall probe mapping experiment), as described in Ref. [49,50]. Although both top and bottom faces of the sample exhibit similar flux distributions profiles, which would suggest a uniformly distributed $J_c$, some underestimation of the true critical current density of the sample by this measurement technique is likely in view of the small, but finite, distance between the Hall probe and the surface of the sample. The theoretical predictions indicate that the threshold field $H_{tr2}$ is, to the first order, proportional to $(J_c)^{1/3}$ [cf. Eq. (8) and (21)], so an increase of $J_c$ by a factor 2 yields an increase in $H_{tr2}$ by a factor ~ 1.26. An underestimation of $J_c$ is therefore not sufficient to explain the observed difference between theory and experiment in this study. The third point is that the model assumes a field-independent $J_c$. In practice, however, all bulk melt-processed materials exhibit some $J_c(B)$ dependence [51] which is sometimes noticeable at field amplitudes of a few tens of mT [52]. In addition, the effects of magnetic field on $J_c$ increase as the sample temperature increases towards the critical temperature $T_c$, which is precisely the case in magneto-thermal experiments described here. The $J_c(B)$ dependence should therefore ideally be taken into account in the model in order to reconcile experimental and theoretical data.

The three arguments outlined above are likely to explain the observed quantitative difference between the measurement and the theoretical predictions. Despite of this difference, the theoretical and the experimental data yield a threshold field and equilibrium temperatures that are of a similar order of magnitude and which display similar features. Note also that our model assumes a homogeneous sample temperature distribution, which is expected to arise when the thermal conductivity $\kappa$ is much larger than the ($U$ $a$) product, such as the dimensionless Biot number $Bi = U\,a\,/\,\kappa$ is much smaller than unity. Using the experimentally determined $U$ et *a* values (cf. Table 1) and assuming a lower bound for the thermal conductivity of YBCO at 77 K ($\kappa \sim$ 1.5 W/m K) [53-57], the Biot number is equal to 0.02. Unlike other studies where non-uniform temperature distributions were observed [58,59], the sample temperature can thus be considered as uniform in the present case, which explains the excellent qualitative agreement between the model and the experiment. This shows that the simple model used here based on the Bean model for an infinite (one-dimensional) geometry might be used to obtain an order of magnitude estimate of heat flow rates and hence to determine whether magneto-thermal effects in a given experimental arrangement are likely to be significant.

*4.4 Two successive field amplitudes*

The following experiment was performed to illustrate the existence of two equilibrium temperatures for a sample characterized by a convective heat transfer coefficient $AU$ smaller than $AU_{lim}$ and subjected to a magnetic field amplitude $H_m$ between $H_{tr1}$ and $H_{tr2}$. The sample was initially cooled to $T_0 = 77.4$ K in a zero field cooling (ZFC procedure) and then subjected to a large magnetic field $H_m = 3.75\ H_{tr2}$ for a given time interval (87 s) in order to allow the thermal steady state to be reached. The corresponding average sample temperature was 91.3 K in this process. The magnetic field amplitude was subsequently lowered to $0.82\ H_{tr2}$. A new steady-state was observed after ~15 s, which corresponded to an average sample temperature equal to 91.12 K.

The application of an AC field amplitude equal to $0.82\ H_{tr2}$ to the sample cooled initially to $T_0$ should yield sample temperature close to 83 K; i.e. the *lower* equilibrium $T_{inf}$, as shown in Fig. 6. In the present case, however, the initial sample temperature is already close to the critical temperature $T_c$. As a result, the equilibrium temperature resulting from the application of $0.82\ H_{tr2}$ is the *upper* equilibrium $T_{sup}$. This is in excellent agreement with the theoretical analysis presented in Section 2(a): two equilibrium temperatures can arise for a given AC field amplitude $H_m$, depending on the sample temperature prior to the application of the AC field. This experiment confirms that the equilibrium temperature of a type-II superconductor subjected to an AC field is not determined in a unique manner but is a function of the initial temperature, which results ultimately from the history of the application of AC fields (i.e. from the magnetic history of the superconductor).

*4.5. Final comments on the role played by the different parameters*

In summary, this study has shown that the central parameter characterizing the importance of magneto-thermal effects is the *upper* threshold field $H_{tr2}$, roughly given by;

$$H_{tr2} \approx \left[ (\gamma) \frac{A U \left(T_c - T_0\right) J_{c0}\, a}{\mu_0 f\, V} \right]^{1/3}, \qquad (38)$$

where the numerical factor $\gamma$ is equal to (3/8) for an infinite slab and equal to (3/16) for an infinite cylinder. The value of $H_{tr2}$ should be as large as possible in order to minimize self-heating effects. This can be achieved by (i) a large heat transfer coefficient $U$, (ii) an initial temperature $T_0$ much smaller than $T_c$, (iii) a large critical current density $J_{c0}$ and (iv) a small frequency. Equation (38) above shows that the sample external area $A$, radius $a$ and volume $V$ combine into a single dimensionless parameter $\zeta = Aa / V$. Although Eq. (38) holds for an *infinite* geometry (i.e. of infinite external area $A$ and volume $V$), it is instructive to express the parameter $\zeta = Aa / V$ for a *finite* cylinder or radius $a$ and height $h$ as follows;

$$\zeta = \frac{A\,a}{V} = 2\left(1 + \frac{a}{h}\right). \qquad (39)$$

A large value of $H_{tr2}$ is therefore achieved for a *small aspect ratio* ($h/a$) of the cylinder. Equation (39) suggests also that two cylindrical samples of different radii but displaying the same aspect ratio should be characterized by identical $H_{tr2}$ values.

Finally, it should be noted that the analytical expressions derived for $H_{tr2}$ provide insight of the *combined* influence of several parameters. As an example, multiplication of the convective heat coefficient $U$ and the frequency $f$ by the same factor is shown to have no effect on the parameter $H_{tr2}$. In addition, due to the 1/3 exponent, none of the parameters appearing in the analytical expressions appears to play a crucial role in the analysis, since doubling any value yields a modification of $H_{tr2}$ by factor of $2^{1/3}$ or $(1/2)^{1/3}$, which is a relative change only of the order of 20-25%.

## 5. Conclusions

We have analyzed both theoretically and experimentally the magneto-thermal effects arising in a bulk high-temperature superconductor subjected to an AC magnetic field. The calculations are based on the critical state model applied to an infinitely long type-II superconducting slab or cylinder, with field-independent $J_c$ and constant heat transfer coefficient $AU$. We have derived analytical expressions for the equilibrium (steady-state) temperature attained by the superconductor. We have defined a limit heat transfer coefficient $AU_{lim}$ and two specific threshold magnetic fields $H_{tr1}$ and $H_{tr2}$. The $AU_{lim}$ value corresponds to the heat transfer coefficient below which two equilibrium temperatures exist in the system. The relevant threshold field to consider is $H_{tr2}$ in the common case where the sample is placed at a constant cryogenic fluid temperature $T_0$ and then subjected to an AC field. Once the applied field amplitude exceeds $H_{tr2}$, the steady-state temperature of the superconductor switches from a "low" equilibrium to a "high" equilibrium temperature. The "low" equilibrium temperature is always smaller than $(T_0 + T_c)/2$ for a slab, and smaller than $(T_0 + 2T_c)/3$ for a cylinder, whereas the "high" equilibrium is close to, but slightly smaller than, the critical temperature $T_c$. At small heat transfer coefficients $AU < AU_{lim}$, a large temperature "window" is forbidden as equilibrium temperature of the superconductor. The theoretical characteristics of the self-heating of a type-II superconductor were confirmed experimentally using magneto-thermal measurements on a bulk melt-processed YBCO sample placed in an AC susceptometer allowing large alternating magnetic fields (~ 80 mT) to be applied. The experimental system is characterized by a reproducible heat transfer coefficient $AU$ that could be determined precisely. The experimental results have shown a correlation between the "high" equilibrium temperature and the amplitude of magnetic field. We have also investigated how the results are modified when the initial sample temperature differs from that of the coolant. We have shown that, in some cases, an AC magnetic field amplitude smaller than the threshold field $H_{tr2}$ can give rise to an equilibrium temperature close to $T_c$. The agreement between the experimental data and the analytical expressions suggests that time-expensive thermal numerical modelling can be replaced conveniently by a simple analytical analysis to yield the right orders of magnitude for simple superconductor geometries and to predict the role played by the experimental parameters on the self-heating characteristics of the superconductor.

**Acknowledgments**

We acknowledge Profs. B. Vanderheyden and R. Cloots for fruitful discussions and comments. We also thank the FNRS, the ULg and the Royal Military Academy (RMA) of Belgium for cryofluid and equipment grants.

Table

Table 1. Numerical value of parameters used for the experimental characterization of the magneto-thermal behaviour of a cylindrical bulk melt-processed YBCO sample.

| Parameter | |
|---|---|
| Critical temperature $T_c$ | 91.6 K |
| Initial temperature $T_0$ | 77.4 K |
| Initial critical current density $J_{c0}$ | $10^3$ A/cm² |
| Sample radius $a$ | 15 mm |
| Sample height $h$ | 12 mm |
| Sample volume $V$ | $8.48 \times 10^{-6}$ m³ |
| Heat transfer coefficient $AU$ | $4.94 \times 10^{-3}$ W/K |
| Convective coefficient $U$ | 1.94 W/m² K |
| Frequency of applied field $f$ | 56 Hz |
| Amplitude of applied field $\mu_0 H_m$ | from 0 to 100 mT |

# Figure 1

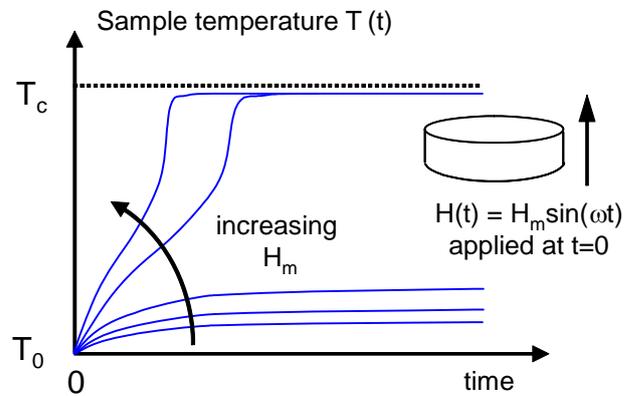

**Figure 1**. Schematic illustration of the self-heating behaviour of a bulk type-II superconductor subjected to an AC magnetic field $H(t) = H_m \sin(\omega t)$, according to the set of experimental data available in the literature. At low field amplitudes $H_m$, the temperature rises up to an equilibrium temperature depending on $H_m$ but being much smaller than the critical temperature $T_c$. On increasing the field amplitude further, a thermal runaway occurs and the sample temperature rises quickly up to an equilibrium temperature that is nearly field-independent and close to $T_c$.

# Figure 2

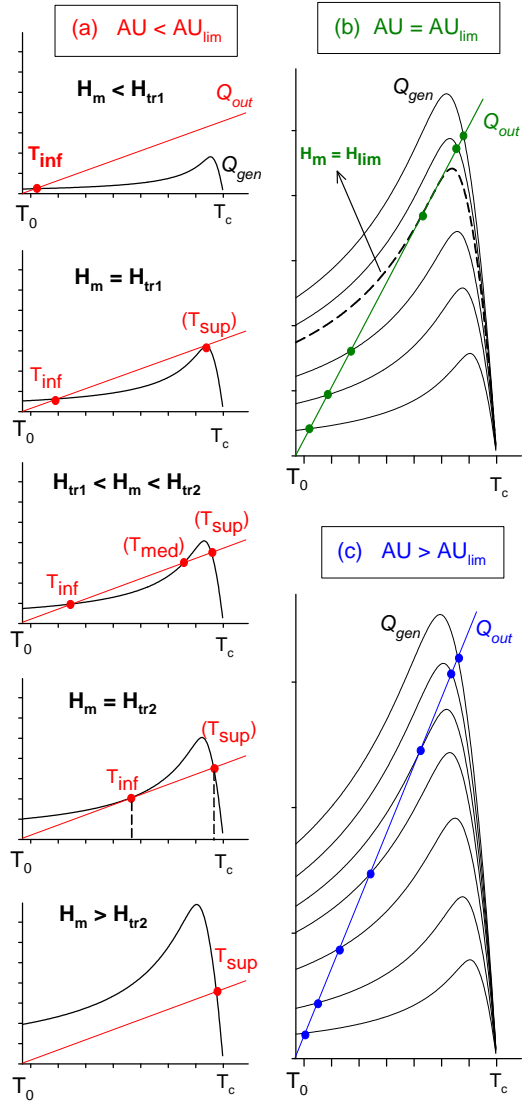

**Figure 2**. Temperature dependence of the AC losses $Q_{gen}(T)$ generated within a type-II superconducting sample placed initially in a cryogenic environment at a temperature $T_0$ and subjected to an AC magnetic field $H(t) = H_m \sin(\omega t)$, compared to the heat flux rate leaving the sample by convection $Q_{out} = AU\,(T - T_0)$. Three different scenarios are investigated, depending on the convective heat transfer coefficient $AU$ with respect to the limit value $AU_{lim}$ (see text). (a) $AU < AU_{lim}$ : from top to bottom, five increasing field amplitudes are considered, corresponding to different numbers of intersections between the $Q_{gen}(T)$ and the $Q_{out}(T)$ curves. (b) Intermediate case ($AU = AU_{lim}$) for which $Q_{gen}(T)$ and $Q_{out}(T)$ intersect at one point and for which the $Q_{gen}(T)$ curve at some amplitude $H_m = H_{lim}$ is tangential to the $Q_{out}(T)$ straight line. (c) $AU > AU_{lim}$: each of the $Q_{gen}(T)$ curves (from top to bottom : increasing field amplitude) intersect the $Q_{out}(T)$ straight line in one single point and none of them is tangential to $Q_{out}(T)$.

# Figure 3

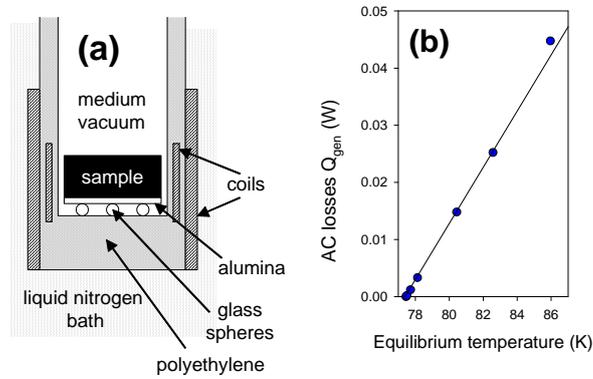

**Figure 3**. (a) Schematic illustration of the bottom of the experimental chamber of the AC susceptometer used for magneto-thermal experiments (b) Experimental data of the AC losses $Q_{gen}$ within sample PL1 (bulk melt-processed YBCO disc, 30 mm diameter and 12 mm thickness) as a function of the equilibrium temperature $T_e$ of the sample. The superconductor is inserted in the sample chamber of the AC susceptometer and its temperature is stabilized at $T_0 = 77.4$ K prior to applying the AC magnetic field.

# Figure 4

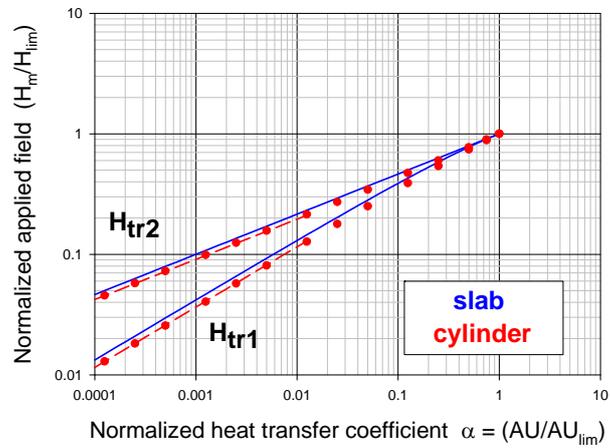

**Figure 4**. Influence of the heat transfer coefficient *AU* on the threshold fields $H_{tr1}$ and $H_{tr2}$ determined analytically for a type-II superconducting slab (plain lines) or numerically for a type-II superconducting cylinder (symbols) whose infinite direction is parallel to the applied AC field. The threshold fields are normalized to the limit field $H_{lim}$; the convective heat transfer coefficient *AU* is normalized to the limit value $AU_{lim}$ (see text). The dashed lines are approximations of the threshold fields for an infinite cylinder that are given by equations (32) and (33).

**Figure 5**

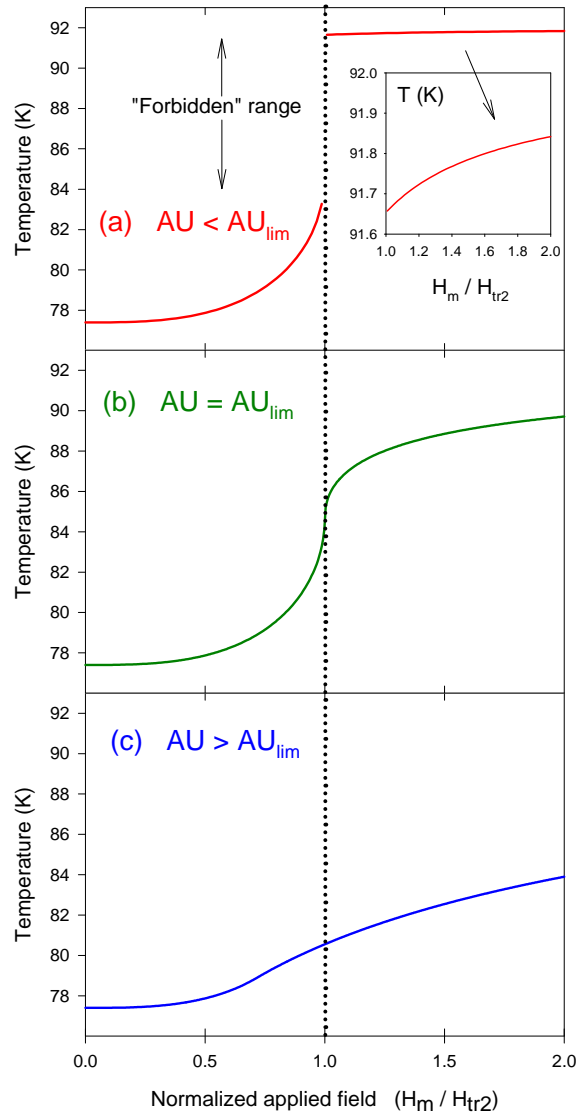

**Figure 5**. Theoretical predictions of the steady-state temperature of a type-II superconducting slab subjected to an AC magnetic field $H(t) = H_m \sin(\omega t)$ parallel to its surface. The initial temperature of the slab before application of the AC field is 77.4 K and the critical temperature $T_c$ is equal to 92 K. The heat transfer coefficient AU is either (a) smaller than (b) equal to or (c) larger than the limit value $AU_{lim}$. The applied field amplitude is normalized to the threshold field $H_{tr2}$. The inset in Fig. 5(a) shows an enlargement of the equilibrium temperature for $H_m > H_{tr2}$.

# Figure 6

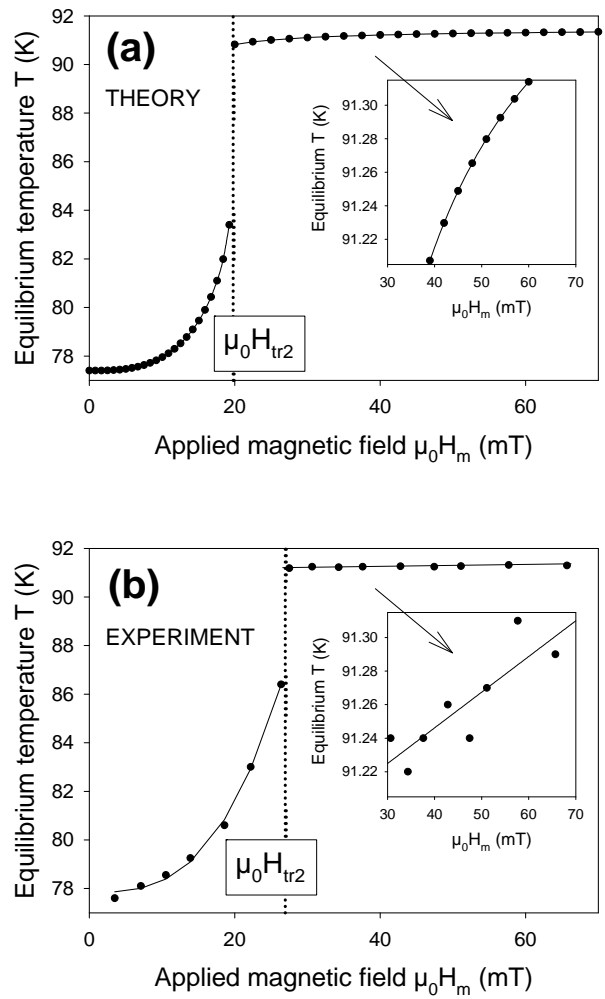

**Figure 6**. (a) Theoretical predictions of the equilibrium (steady-state) temperature as a function of the AC field amplitude for an infinite type-II superconducting cylinder; the parameters are those listed in Table 1. (b) Measured equilibrium temperature of sample PL1 subjected to an AC magnetic field $H(t) = H_m \sin(\omega t)$. For both plots, the vertical line shows the experimental threshold field $H_{tr2}$. The inset shows the equilibrium temperature for $H_m > H_{tr2}$.